\documentclass[manuscript,screen,nonacm]{acmart}

\usepackage{algorithm}
\usepackage{algpseudocode}
\usepackage{graphicx}

\usepackage{comment}
\usepackage{float}

\begin{document}
\title{Economic Dynamics of Agents}

\author{Dibakar Das}
\orcid{0000-0002-6731-8975}
\email{dibakard@acm.org}
\affiliation{%
  \institution{IIIT Bangalore}
  \state{Karnataka}
  \country{India}
  \postcode{560100}
}

\begin{abstract}

  Post-pandemic world has thrown up several challenges, such as, high inflation, low growth, high debt, collapse of economies, political instability, job losses, lowering of income in addition to damages caused natural disasters, more convincing attributed to climate change, apart from existing inequalities. Efforts are being made to mitigate these challenges at various levels. To the best of the knowledge of the author, most of the prior researches have focussed on specific scenarios, use cases, inter-relationships between couple of sectors and more so on optimal policies, such as, impact of carbon tax on individuals, interaction between taxes and welfare, etc. However, not much effort have been made to understand the actual impact on individual agents due to diverse policy changes and how agents cope with changing economic dynamics. This paper considers progressive deteriorating conditions of increase in expense, degrading environmental utility, increase in taxation, decrease in welfare and lowering of income with recourse to inherited properties, credits and return on investments, and tries to understand how the agents cope with the changing situations using an agent based model with matrices related to savings, credits, assets. Results indicate that collapse of agents' economic conditions can be quite fast, sudden and drastic for all income groups in most cases.
\end{abstract}


\maketitle

\section{Introduction}
Several economic challenges have cropped up (more aggressively) following the subsiding of the global pandemic. One of the major concerns has been high inflation \cite{cite_long_term_impact_of_pandemic_on_inflation} leading of increase in household expenses. With low economic growth there are job losses leading to lowering of incomes of individuals \cite{cite_gender_inequality_in_job_loss_pandemic}. Several smaller economies are on the verse of collapse due to high debt leading to political instabilities. Added to these is the economic impact of natural disasters more convincingly being attributed to climate change. Though, many ways of public welfare and other economic measures are being taken to mitigate the deteriorating conditions, individuals are trying their best to navigate through these adverse situations with their available options,e.g, credits, savings, etc., and sometimes conflicting impact of policies. Hence, it is of paramount importance to understand how individual agents deal with these economic challenges with computational models.

Interaction of different topics in relationship to economics have been studied over decades. Some of the works have concentrated on economic consequences from climate change \cite{cite_new_ecological_socio_economic_divide_europe}\cite{cite_effective_climate_policies_amid_conflicting_information}\cite{cite_social_cost_of_carbon_with_ethical_values}. A number of works have focussed on sustainable development as a the way forward \cite{cite_from_welfare_and_growth_to_degrowth_and_sustainable_welfare}\cite{cite_support_for_sustainability_in_europe_with_welfare_and_climate_policies}. Impact of climate change on different income groups have also been investigated \cite{cite_climate_policies_impact_on_income_groups}\cite{cite_impact_of_climate_policies_on_labour_and_income}.
Climate change impact on food security has been studied in \cite{cite_climate_change_and_food_security}. A consumption based tax systems apart from income tax based has been suggested for better welfare measures \cite{cite_income_tax_vs_consumption_tax_and_welfare}. Joint solutions to challenges of economic growth and income equality has been opined in
\cite{cite_target_both_income_inequality_and_economic_growth_oecd}. Benefits of earned income tax credits particularly to the low income groups has been highlighted in \cite{cite_earned_income_tax_credit_builds_hopes}. Different aspects of carbon tax, such as, public opinion \cite{cite_opposition_to_carbon_tax_europe} and recirculation of such collected taxes have been studied \cite{cite_carbon_dividend_from_carbon_tax_for_income_inequality}. Most of the research have concentrated on policy aspects and their empirical studies. However, agents' (individuals') experiences to both policies which can themselves be of conflicting nature as well as their prevailing economic conditions. Study of these experiences from an integrated and holistic perspective using a computational model have been largely unexplored.

The paper considers five increasingly deteriorating scenarios for agents over time: 1) increase in expense, 2) increase in expense and decrease in environment utility, 3) increase in expense, decrease in environment utility and rise in taxes, 4) increase in expense, decrease in environment utility, rise in taxes and lowering of public welfare, 5) increase in expense, decrease in environment utility, rise in taxes, lowering of public welfare and reduced income. An agent based model is developed where agents are divided into three income groups, high, middle and low. Each agent has inputs, such as, environment utility, income, public welfare, inherited assets. Agents also have access to credits, saves some amount which they either partly invest and derive return on investments (ROI) or prefer to keep those for future use. Each agent pays taxes and has its expense as required. Each of the variables, such as, income, inherited assets, etc., are assumed to be drawn from uniform distribution within ranges based on their income groups. The environment utility is defined as a decreasing hyperbolic function as time progresses. To avoid complexity, increase or decrease in other parameters are assumed to be linear (explained in detail in section \ref{section_system_model}). As the agents navigate through the above scenarios over time, several parameters relating to their overall economic health in terms of their saving, credits, inherited assets and economic stress is measured through simulations. Results show the transitions occur much earlier in time as economic conditions deteriorates for the agents and they can be very sudden with the lower income groups getting affected first, though it is matter of time before other two groups also show sharp decline. To the best of the knowledge of the authors, none of the previous works has looked into these scenarios deteriorating with an integrated computational model.
\section{Literature Survey}
Several research works have published addressing different scenarios of interaction between economics and other multiple factors. This section discusses some these well cited works focussing more on the ones in the past decade.

Climate change is leading to major economic consequences. \cite{cite_new_ecological_socio_economic_divide_europe} opines that some of the welfare measures may have to traded off in developed welfare states (in Europe) to pay for the cost of climate change and may bring an additional detrimental ecological aspect to already existing socioeconomic divide. 
Need for effective policies and optimal utilization of resources to deal with climate change rather than misdirected ones leading to contradictory goals has been highlighted in \cite{cite_effective_climate_policies_amid_conflicting_information}. The author analyses various reports, finds out contradictions and opines that the steps taken to deal with climate change are ineffective.
Comprehensive guidelines to design fiscal policies in nations across the world to deal with development and climate change has been put forward in \cite{cite_worldbank_fiscal_policies_development_and_climate_change}. \cite{cite_from_welfare_and_growth_to_degrowth_and_sustainable_welfare} proposes transformation of existing welfare state policy towards a more sustainable approach rather than looking into economic growth only. The paper underscores the for holistic approach to deal with the contractions in social welfare, economic growth and sustainability.
A new methodology to evaluate social cost of carbon using ethical values rather than utility based approach has been proposed in \cite{cite_social_cost_of_carbon_with_ethical_values}.
A detailed study of attitude of people in Europe towards climate and welfare policies have been studied in \cite{cite_support_for_sustainability_in_europe_with_welfare_and_climate_policies}. Authors studied the impact of demography, education, social, political and income on individual attitudes. They find that socio-democratic countries and people with higher education levels  and moderate views seems to support welfare and climate policies adhering to sustainability.
In this very interesting work \cite{cite_climate_policies_impact_on_income_groups}, authors underscore the importance of designing climate policies without impacting different income groups adversely.
\cite{cite_impact_of_climate_policies_on_labour_and_income} studies how green policies can impact labour market and wage distribution in different sectors with five worker categories.
Authors make detailed studies of hundreds of assessments of climate and energy policies over last couple of decades on societal impact in terms of electricity access, energy affordability, employment, livelihoods and poverty, etc., in \cite{cite_analysis_of_policies_over_decades_most_donot_deliver_goals}. They find that most of the cases the policies do not deliver the required social outcomes though a limited number of them do serve the purposes of both climate and society.
A suggestion to move away from growth oriented welfare for sustainability and accepting a degrowth path has been put forward in \cite{cite_move_away_from_growth_to_sustaianable_welfare_max_and_basic_income}. This paper empirically studies the impact of having maximum and basic incomes, taxing wealth and meat, and reduction in working hours as means to achieve sustainable welfare.
\cite{cite_social_cost_of_carbon} estimates the social cost of carbon and its relationship to income elasticity across different countries and finds that large and poor ones have high estimated costs.
The need for appropriate fiscal policies to deal with the costs due to climate change in terms of  preventive and remedial measures while maintaining economic growth has been studied in \cite{cite_fiscal_polity_for_climate_change}.
Author considers different criteria and types of welfare under the ambit of universal basic income, universal basic service and universal basic vouchers as ways to achieve sustainable welfare in \cite{cite_sustainable_welfare_ubi_ubs_ubv_6_criteria_9_types_of_welfare}. The need to consider  environmental externalities to decide on entrepreneurial successes has been highlighted in \cite{cite_entreprise_success_and_env_externalities}. \cite{cite_climate_change_and_food_security} highlights the need of efficient policies for food security in the context of climate change. Principles for green economic growth for sustainable development, efficient resource utilization and reduction of environmental risks have been proposed in \cite{cite_green_economic_principles_for_sustainability}.
In the working paper \cite{cite_impact_of_climate_change_policies_on_macroeconomics}, authors study aggressive climate change mitigation policies can lead to challenges in macroeconomic factors.
Authors study whether climate changes can have regressive effect on different economic sections of the society in \cite{cite_climate_change_policies_may_be_regressive_to_different_income_groups}. A study of how welfare and climate policies can complement each other has been studied in \cite{cite_welfare_and_climate_policies_together_go_well}.
\cite{cite_income_tax_vs_consumption_tax_and_welfare} studies the impact of consumption taxes instead of taxing incomes with low taxes on essentials and high taxes non-essentials. The study finds that progressive income taxes and low consumption taxes can yield better welfare measures.

Economic Inequality has always been a major concern over the world. \cite{cite_oecd_taxes_transfer_to_takle_inequality} studies how taxes and transfer of welfare measures can help tackling inequality.
\cite{cite_ireland_2008_crisis_impact} studies the impact of economic crisis (in Ireland) and finds that the top and the bottom  deciles are most impacted. Though, welfare policies helped the bottom decile to lower their losses, other income groups had to deal with higher losses due to impact of the recession.
Policies to effectively target two challenges of income inequality and achieving higher economic growth simultaneously has been studied in \cite{cite_target_both_income_inequality_and_economic_growth_oecd}.
Study on how earned income tax credit and child tax credit helps low and middle income groups in their economic well-being has been described in \cite{cite_income_child_taxes_credit_low_and_middle}. Also, in \cite{cite_earned_income_tax_credit_builds_hopes} authors uses surveys to show how earned income tax credits provide resources to cover the financial gaps of low income groups and help them build goals for assets. \cite{cite_usa_credit_and_asset_bubble_impact_on_debt} analyses the impact of credit and asset bubble from 2001-2007 on debt burden and financial insecurities of American household, especially young and senior citizens.

Carbon tax is often considered a viable option to deal with climate change. Major research have focussed on carbon pricing, opposition to such taxes and recycling of revenue back to society. A machine learning based approach to study the reasons behind opposition to carbon tax has been studied in \cite{cite_opposition_to_carbon_tax_europe}. Author finds that personal responsibility, public trust and recycling of carbon taxes to households are key reasons behind such opposing trends in Europe. Author highlights the need for differential and dynamic approach to carbon pricing to deal with climate change in \cite{cite_joseph_stiglitz_differential_pricing}. \cite{cite_political_support_for_policies_beyond_ecomomics} highlights the need for political support and appropriate usage of revenues under evolving political and economic situations as against only relying on economic principles for carbon pricing. A proposal to use carbon tax to fund carbon dividend as a way to deal with income inequality has been studied in \cite{cite_carbon_dividend_from_carbon_tax_for_income_inequality}.

From the above discussion it is evident that most papers focus on top down macro policies and try to understand their impact empirically. Many of these policies deal with specific scenarios, e.g., climate and food security, carbon tax and distribution of revenues back to society, economic growth and income inequality, etc. Most of the prior work do not consider the situation for example, when income and welfare schemes are eaten away by higher expenses of agents which could be because of a natural disaster. This paper tries to understand using a computational model the experiences of agents from multiple policies and prevailing deteriorating economic circumstances as they cope using their own individual local policies and available resources.
\section{System Model}\label{section_system_model}
\begin{figure}[H]
\centering
\includegraphics[width=\columnwidth]{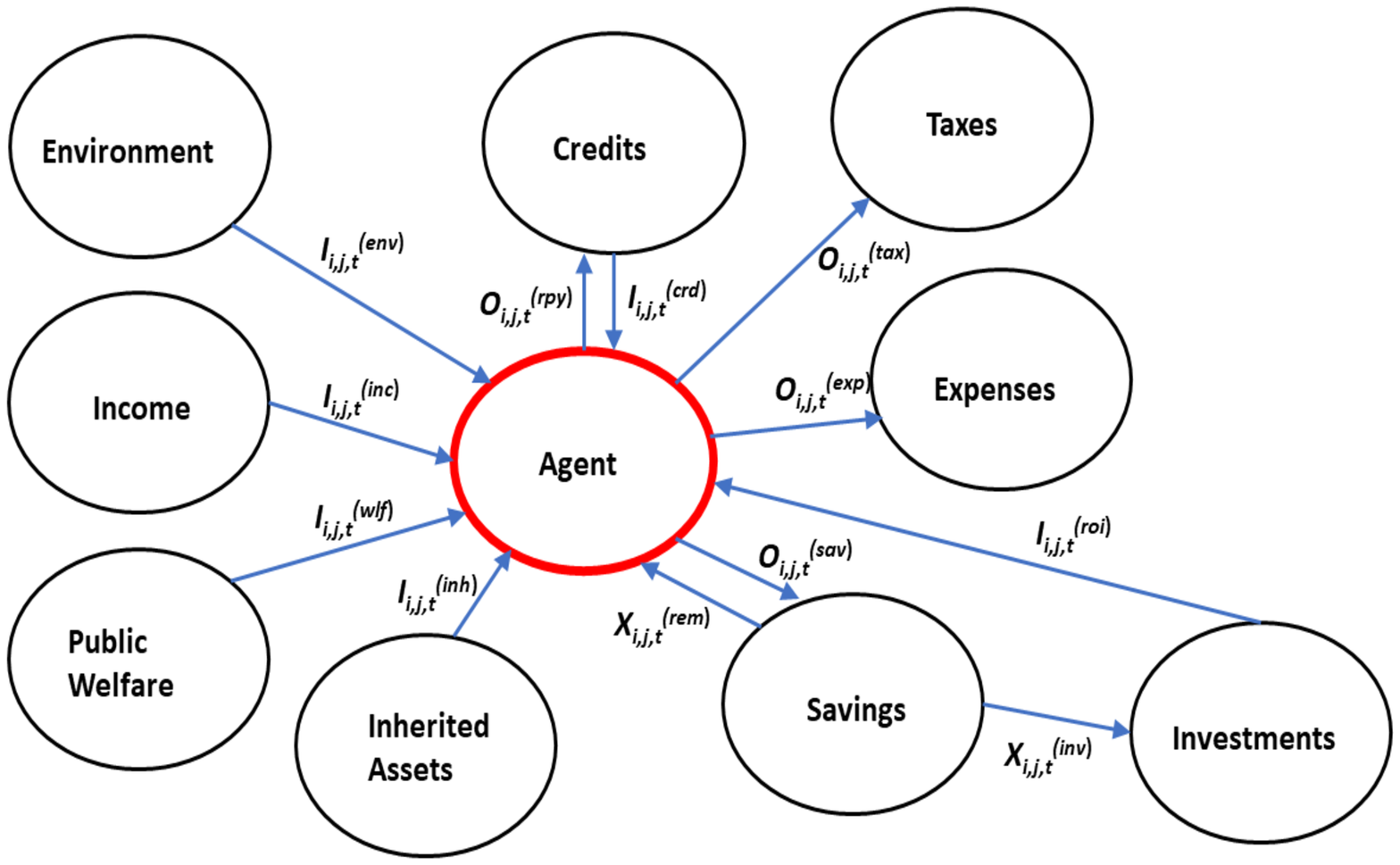}
\caption{Economic interaction of an agent}
\label{fig_block_diagram_inkscape}
\end{figure}

Fig. \ref{fig_block_diagram_inkscape} shows the interaction of an agent with its economic environment. All incoming arrows are its available sources of wealth (inputs) and all outgoing ones are the areas where the wealth goes (outputs). An agent tries to maintain a balance among the inputs and outputs under changing scenarios (explained latter). Let there be $N$ agents in the system. These $N$ agents belong to $G$ income groups. $I_{i,j,t}^{(env)}$ denotes the input utility of $i^{th}$ agent from the $j^{th}$ income group derives from the environment at time $t$. $I_{i,j,t}^{(inc)}$ is the agent's direct income, e.g., salary. $I_{i,j,t}^{(wlf)}$ is the public welfare it is entitled to. The agent inherits certain amount of wealth $I_{i,j,t}^{(inh)}$ from its legacy. The agent also gets a return on its investments amounting to $I_{i,j,t}^{(roi)}$. In case of need, it may borrow money from institutions in form of credits, insurances, etc., an amount of $I_{i,j,t}^{(crd)}$ and repays those credits, premiums, etc., which amount to $O_{i,j,t}^{(rpy)}$. The agent also pays direct tax of amount $O_{i,j,t}^{(tax)}$ on its income $I_{i,j,t}^{(inc)}$. Its expense (which may include indirect taxes) is $O_{i,j,t}^{(exp)}$. The agent also saves an amount of $O_{i,j,t}^{(sav)}$. In case of need, it also withdraws from this savings an amount of $X_{i,j,t}^{(rem)}$. It also invests an amount of $X_{i,j,t}^{(inv)}$ from its savings.

Income $I_{i,j,t}^{(inc)}$, is a uniform random value $X_{i,j,t}^{(inc)}$ between $I_j^{(min)}$ and $I_j^{(max)}$. Each income group $j$ has separate expense rate $\gamma_j$ and tax rate $\delta_j$. Public welfare  $I_{i,j,t}^{(wlf)}$, is a uniform random value $X^{(wlf)}_{i,j,t}$ between $W_j^{(min)}$ and $W_j^{(max)}$. $I^{(inh)}_{i,j,0}$ is also initialised to a uniform random value between $V_j^{(min)}$ and $V_j^{(max)}$. The following sections describe changing scenarios an agent experiences along with the models for variation of the above parameters.
\subsection{Increase in expense over time}
Expenses of agents can increase due to various reasons, such as, increased cost of living, inflation, etc. Increase in expense of $i^{th}$ agent over the time $t = 1,2,..,T$ is defined by equation (\ref{eqn_out_exp}). Expense of each agent $i$ is assumed to increase linearly with time $t$.
\begin{equation}\label{eqn_out_exp}
O^{(exp)}_{i,j,t} = [1 + \alpha^{(exp)}_{i,j} t] \gamma_j I^{(inc)}_{i,j,t}
\end{equation}
where $\alpha^{(exp)}_{i,j}$, $0 < \alpha^{(exp)}_{i,j} \le 1$, is used to control the increase over time.
\subsection{Decrease in environment utility over time}
Each agent derives a certain utility from environment, such as, natural resources, suitable climatic conditions, etc. To model the varying environment utility, hyperbolic sine function, i.e., sinh(), is used. Hyperbolic sine function is used because environmental utility can be modeled from a situation when it provides the highest $A_{i,j}$ for the $i^{th}$ belonging to the $j^{th}$ group then progressively going down to zero and further declining (where degrading environment itself creates negative impact, e.g., extreme climatic conditions, natural disasters, etc.) to a minimum value of $-A_{i,j}$. Environment utility of $i^{th}$ agent belonging to the $j^{th}$ group at time $t$, $I^{(env)}_{i,j,t}$, varies as in equation (\ref{eqn_input_env}) and it is normalized between $A_{i,j}$ and $-A_{i,j}$.
\begin{equation}\label{eqn_input_env}
I^{(env)}_{i,j,t} = [(A_{i,j} - (-A_{i,j})]\frac{\sinh(\alpha^{(env)}_{i,j} t - \frac{T}{2}) - min{(\{\sinh(\alpha^{(env)}_{i,j} t - \frac{T}{2})\})}}{max{(\{\sinh(\alpha^{(env)}_{i,j} t - \frac{T}{2} )\})}-min{(\{\sinh(\alpha^{(env)}_{i,j} t - \frac{T}{2})\})}} + (-A_{i,j})
\end{equation}
$I^{(env)}_{i,j,t}$ is generated for the period $t = 1,2,..,T$ and $\alpha^{(env)}_{i,j}$, $0 < \alpha^{(env)}_{i,j} \le 1$, is the control parameter.
\subsection{Increase in taxes over time}
Each income group has separate tax rate. Taxes is defined to increase linearly as shown in equation (\ref{eqn_output_tax}).
\begin{equation}\label{eqn_output_tax}
O^{(tax)}_{i,j,t} = [1 + \alpha^{(tax)}_{i,j}\frac{t}{T}]\delta_j I^{(inc)}_{i,j,t}
\end{equation}
where $\delta_j$ is the tax rate for the $j^{th}$ income group. As before, $\alpha^{(tax)}_{i,j}$, $0 < \alpha^{(tax)}_{i,j} \le 1$, is used to adjust the rate of increase of taxes.
\subsection{Decrease in public welfare over time}
Each agent $i$ in income group $j$ has separate rate $\alpha^{(wlf)}_{i,j}$ to receive public welfare. Decreasing public welfare received by an agent is defined by equation (\ref{eqn_output_welfare}).
\begin{equation}\label{eqn_output_welfare}
I^{(wlf)}_{i,j,t} = [1 - \alpha^{(wlf)}_{i,j}\frac{t}{T}]X^{(wlf)}_{i,j,t},
\end{equation}
where $0 < \alpha^{(wlf)}_{i,j} \le 1$ and $[1 - \alpha^{(wlf)}_{i,j}\frac{t}{T}] > 0$.
\subsection{Decrease in income over time}
Finally, income can decrease due to job losses and other factors which is defined by equation (\ref{eqn_output_income}).
\begin{equation}\label{eqn_output_income}
I^{(inc)}_{i,j,t} = [1 - \alpha^{(inc)}_{i,j}\frac{t}{T}]X^{(inc)}_{i,j,t}
\end{equation}
where $\alpha^{(inc)}_{i,j}$, $0 < \alpha^{(inc)}_{i,j} \le 1$, is the control parameter and $[1 - \alpha^{(inc)}_{i,j}\frac{t}{T}] > 0$.

Thus, savings of agents after taxes and expenses is given by equation (\ref{eqn_output_save}).
\begin{equation}\label{eqn_output_save}
O^{(sav)}_{i,j,t} = I^{(inc)}_{i,j,t} - O^{(tax)}_{i,j,t} - O^{(exp)}_{i,j,t}
\end{equation}
If agent has a total accumulated credit of $I^{(crd)}_{i,j,t-1}$ at time $t-1$ and repays $\eta^{(rpy)}_i I^{(crd)}_{i,j,t-1}$ at time $t$ then credit repayment $O^{(rpy)}_{i,j,t}$ has to be subtracted from $O^{(sav)}_{i,j,t}$. Hence,
\begin{equation}\label{eqn_input_credit_output_repay}
O^{(rpy)}_{i,j,t} =  \eta^{(rpy)}_i I^{(crd)}_{i,j,t-1}
\end{equation}
\begin{equation}\label{eqn_output_save_after_repay}
O^{(sav)}_{i,j,t} = O^{(sav)}_{i,j,t} - O^{(rpy)}_{i,j,t}
\end{equation}
Maximum credit limit $I^{(crd)}_{i,j,0}$ is initialized to $\mu^{(crd)}_i I^{(inc)}_{i,j,0}$ at the beginning, i.e., at $t=0$.
\begin{equation}\label{eqn_output_save_after_repay}
I^{(crd)}_{i,j,0} = \mu^{(crd)}_i I^{(inc)}_{i,j,0}
\end{equation}
Investments is only possible if $O^{(sav)}_{i,j,t} > 0$.
If $\rho^{(inv)}_i$, $0 < \rho^{(inv)}_i \le 1$, is the rate from saving invested then this amount is given by
\begin{equation}
X^{(inv)}_i = \rho^{(inv)}_i O^{(sav)}_{i,j,t}
\end{equation}
and ROI is given by
\begin{equation}
I^{(roi)}_{i,j,t} = I^{(roi)}_{i,j,t-1} + \rho^{(roi)}_i X^{(inv)}_i
\end{equation}
where $\rho^{(roi)}_i$, $0 < \rho^{(inv)}_i \le 1$ is the rate of ROI.
Remaining savings after investment of $i^{th}$ agent is given by
\begin{equation}
X^{(rem)}_{i,j,t} = X^{(rem)}_{i,j,t-1} + [1- \rho^{(inv)}_i] O^{(sav)}_{i,j,t}
\end{equation}
Net balance of $i^{th}$ agent is defined as
\begin{equation}\label{eqn_net_output_save}
X^{(net)}_{i,j,t} = X^{(net)}_{i,j,t-1} + I^{(env)}_{i,j,t} + I^{(wlf)}_{i,j,t} + I^{(roi)}_{i,j,t} + X^{(rem)}_{i,j,t} - abs[O^{(sav)}_{i,j,t}].
\end{equation}
Absolute value of $O^{(sav)}_{i,j,t}$ is used to take care of it being negative when expenses and taxes together are more than the income.
\subsection{Policy of agents}
\begin{algorithm}
\scriptsize
\caption{Balancing policy of agents}\label{algo_agent_balance_policy}
\begin{algorithmic}[1]
\Procedure{AgentPolicy()}{}
\\
\\\hspace{5mm}/* If net balance < 0 then apply following policy */
\\\hspace{5mm}$If (X^{(net)}_{i,j,t} < 0)$
\\\hspace{10mm}/* if agent has extra saving after then adjust from that amount */
\\\hspace{10mm}$If (X^{(rem)}_{i,j,t} > 0)$
\\\hspace{15mm}$deficit = abs(X^{(net)}_{i,j,t})$
\\\hspace{15mm}$If (X^{(rem)}_{i,j,t} > deficit)$
\\\hspace{20mm}$X^{(rem)}_{i,j,t} = X^{(rem)}_{i,j,t} - deficit$
\\\hspace{20mm}$X^{(net)}_{i,j,t} = X^{(net)}_{i,j,t} + deficit$
\\\hspace{15mm}$else$
\\\hspace{20mm}$X^{(net)}_{i,j,t} = X^{(net)}_{i,j,t} + X^{(rem)}_{i,j,t}$
\\\hspace{20mm} $X^{(rem)}_{i,j,t} = 0$
\\\hspace{15mm}$endif$
\\\hspace{10mm}$endif$
\\
\\\hspace{10mm}/* Still, if net balance < 0 then adjust from ROI
\\\hspace{10mm}$If (X^{(net)}_{i,j,t} < 0)$
\\\hspace{15mm}$deficit = abs(X^{(net)}_{i,j,t})$
\\\hspace{15mm}$If (I^{(roi)}_{i,j,t} > deficit)$
\\\hspace{20mm}$I^{(roi)}_{i,j,t} = I^{(roi)}_{i,j,t} - deficit$
\\\hspace{20mm}$X^{(net)}_{i,j,t} = X^{(net)}_{i,j,t} + deficit$
\\\hspace{15mm}$else$
\\\hspace{20mm}$X^{(net)}_{i,j,t} = X^{(net)}_{i,j,t} + I^{(roi)}_{i,j,t}$
\\\hspace{20mm}$I^{(roi)}_{i,j,t} = 0$
\\\hspace{15mm}$endif$
\\\hspace{10mm}$endif$
\\
\\\hspace{10mm}/* Still, if net balance < 0 then either adjust from credits or inherited assets */
\\\hspace{10mm}$If (X^{(net)}_{i,j,t} < 0)$
\\\hspace{15mm}/* Agent with index even prefers to avail credit */
\\\hspace{15mm}$If (i\ is\ even)$
\\\hspace{20mm}$deficit = abs(X^{(net)}_{i,j,t})$
\\\hspace{20mm}$If (I^{(crd)}_{i,j,t} > deficit)$
\\\hspace{25mm}$I^{(crd)}_{i,j,t} = I^{(crd)}_{i,j,t} - deficit$
\\\hspace{25mm}$X^{(net)}_{i,j,t} = X^{(net)}_{i,j,t} + deficit$
\\\hspace{20mm}$else$
\\\hspace{25mm}$X^{(net)}_{i,j,t} = X^{(net)}_{i,j,t} + I^{(cred)}_{i,j,t}$
\\\hspace{25mm}$I^{(crd)}_{i,j,t} = 0$
\\\hspace{20mm}$endif$
\\\hspace{15mm}/* Agent with index odd prefers to avail inherited assets */
\\\hspace{15mm}$else$
\\\hspace{20mm}$deficit = abs(X^{(net)}_{i,j,t})$
\\\hspace{20mm}$If (I^{(inh)}_{i,j,t} > deficit)$
\\\hspace{25mm}$I^{(inh)}_{i,j,t} = I^{(inh)}_{i,j,t} - deficit$
\\\hspace{25mm}$X^{(net)}_{i,j,t} = X^{(net)}_{i,j,t} + deficit$
\\\hspace{20mm}$else$
\\\hspace{25mm}$X^{(net)}_{i,j,t} = X^{(net)}_{i,j,t} + I^{(inh)}_{i,j,t}$
\\\hspace{25mm}$I^{(inh)}_{i,j,t} = 0$
\\\hspace{20mm}$endif$
\\\hspace{15mm}$endif$
\\\hspace{10mm}$endif$
\\\hspace{5mm}$endif$

\EndProcedure
\end{algorithmic}
\end{algorithm}
The agents apply the policies in \textbf{Algorithm \ref{algo_agent_balance_policy}} when $X^{(net)}_{i,j,t} \ge 0$ to stabilize their economic conditions.
As the first attempt, agents try to use their remaining savings $X^{(rem)}_{i,j,t}$ in lines 6-15. If $X^{(net)}_{i,j,t} < 0$ still, then they try to use their ROI $I^{(roi)}_{i,j,t}$ in lines 18-27. Even then if $X^{(net)}_{i,j,t} < 0$ then half of the agents (even ones) go for credit options in lines lines 32-40 and the remaining half (odd ones) use their inherited assets in lines 42-51. After applying these policies, agents with $X^{(net)}_{i,j,t} \ge 0$ are those who could balance inputs and outputs and stay afloat. Those agents who still have $X^{(net)}_{i,j,t} < 0$ are designated as \emph{stress agents}.
\subsection{Matrices}
The emergent behaviour of the following parameters over time are considered and studied in this paper.
\subsubsection{Net Balance Amount}
Net Balance Amount (NBA) at time $t$ is the sum of $X^{(net)}_{i,j,t}$ for all agents. Increase in this value means better economic condition and vice versa.
\begin{equation}
NBA_t = \sum_{i=1}^N X^{(net)}_{i,j,t}
\end{equation}
\subsubsection{Remaining Saving Left}
Remaining Saving Left (RSL) at time $t$ is the sum of $X^{(rem)}_{i,j,t}$ for all the agents. Increase in this value also means improved economic condition and vice versa.
\begin{equation}
RSL_t = \sum_{i=1}^N X^{(rem)}_{i,j,t}
\end{equation}
\subsubsection{Number of Stressed Agents}
Number of Stressed Agents (NSA) at time $t$ is the cardinality of the set of agents with $X^{(net)}_{i,j,t} < 0$ after applying the policy algorithm. Increase in this value means higher economic distress.
\begin{equation}
NSA_t = \|\{i: X^{(net)}_{i,j,t} < 0\}\|
\end{equation}
\subsubsection{Return on Investment}
Return on Investment (ROI) at time $t$ is the sum of $I^{(roi)}_{i,j,t}$ of all the agents. Increase in this value also means economic improvement and vice versa.
\begin{equation}
ROI_t = \sum_{i=1}^N I^{(roi)}_{i,j,t}
\end{equation}
\subsubsection{Remaining Credit Limit}
Remaining Credit Limit (RCL) at time $t$ is the sum of $I^{(crd)}_{i,j,t}$ for all the agents. Drop in this value means economic distress and vice versa.
\begin{equation}
RCL_t = \sum_{i=1}^N I^{(crd)}_{i,j,t}
\end{equation}
\subsubsection{Remaining Inherited Assents}
Remaining Inherited Assents (RIA) at time $t$ is the sum of $I^{(inh)}_{i,j,t}$ for all the agents. Decrease in this value also means economic distress.
\begin{equation}
RIA_t = \sum_{i=1}^N I^{(inh)}_{i,j,t}
\end{equation}
\subsubsection{Total Stressed Credits}
Total Stressed Credits (TSC) is difference between credits taken and amount repaid. If an agent has maximum credit limit $I^{(crd)}_{i,j,0}$ of 10 and has taken loan of 6, it has remaining credit $I^{(crd)}_{i,j,t}$ of 4. If it repays 2 then stressed credit is (6-2) = 4. Increase in this value also means deteriorating economic condition.
\begin{equation}
TSC_t = \sum_{i=1}^N I^{(crd)}_{i,j,0} - I^{(crd)}_{i,j,t} -  O^{(rpy)}_{i,j,t}
\end{equation}
\subsubsection{Savings after all Expenses}
Savings after all Expenses (SAE) at time $t$ is the sum of $O^{(sav)}_{i,j,t}$ after paying taxes, expenses and credit repayment for all agents. Decrease in this value means deteriorating economic condition.
\begin{equation}
SAE_t = \sum_{i=1}^N O^{(sav)}_{i,j,t}
\end{equation}
\subsubsection{Number of Stressed Agents in income group $j$}
Number of Stressed Agents (NSA) in income group $j$ at time $t$ is the cardinality of the set of agents $X^{(net)}_{i,j,t} < 0$. Increase in this value also means higher economic distress.
\begin{equation}
NSA_{j,t} = \|\{i: X^{(net)}_{i,j,t} < 0, j \in \{1,2,..,G\} \}\|
\end{equation}
\section{Results}\label{ection_results}
This section presents the results from numerical evaluation of the model described above. Following scenarios are considered as explained above.
\begin{enumerate}
\item Impact of increase in expense of each agent
\item Impact of increase in expense and decrease in environment utility of each agent
\item Impact of increase in expense, decrease in environment utility and increase in taxation of each agent
\item Impact of increase in expense, decrease in environment utility, increase in taxation and decrease in welfare utility of each agent
\item Impact of increase in expense, decrease in environment utility, increase in taxation, decrease in welfare utility and decrease in income of each agent
\end{enumerate}

In all the following results, the rules listed below hold good.
\begin{enumerate}
\item   All the parameters are normalized to different ranges for better readability.
\item   NBA, RSL , ROI,  RCL, RIA, SAE, NSA and TSC are depicted with red, black, orange, brown, blue, pink, green and cyan lines respectively.
\item The normalization ranges of the above parameters are listed below.
    \begin{itemize}
        \item NBA - (0.0 - 1.0)
        \item RSL - (0.0 - 0.9)
        \item NSA - (0.0 - 0.8)
        \item ROI - (0.0 - 0.7)
        \item RCL - (0.0 - 0.6)
        \item RIA - (0.0 - 0.5)
        \item TSC - (0.0 - 0.4)
        \item SAE - (0.0 - 0.3)
    \end{itemize}
\item The \emph{x}-axis always represents the time steps, i.e., simulation iterations.
\item The \emph{y}-axis always represents the normalized values of the parameters.
\end{enumerate}
Values of the model parameters used for numerical evaluation are are presented in Table \ref{table_model_parameters}.
\begin{table}[H]
\scriptsize
  \caption{Simulation Parameters}
  \centering
  \begin{tabular}{|p{3cm}|p{1.5cm}|p{2.5cm}|}
  \hline
  Parameter & Value & Description\\  [0.5ex]
  \hline
  $N$ & 100 & Number of agents\\
  \hline
  $G$ & 3 & Number of income groups: High $\rightarrow$ 10\% of agents, Medium $\rightarrow$ 30\% of agents and Low $\rightarrow$ 60\% of agents\\
  \hline
  $\delta_{1}, \delta_{2}, \delta_{3}$ & 35\%, 20\%, 20\% & Tax rates - High $\rightarrow$ 35\%, Medium $\rightarrow$ 20\%,  Low $\rightarrow$ 10\%  \\
  \hline
  $\gamma_{1}, \gamma_{2}, \gamma_{3}$ & 25\%, 40\%, 98\% & Expense rates - High $\rightarrow$ 25\%, Medium $\rightarrow$ 40\%,  Low $\rightarrow$ 98\%  \\
  \hline
  $I_1^{(min)}$, $I_2^{(min)}$, $I_3^{(min)}$  & 8, 4, 1 & Minimum income of three income groups\\
  \hline
  $I_1^{(max)}$, $I_2^{(max)}$, $I_3^{(max)}$  & 10, 7, 3 & Maximum income of three income groups, Income of an agent is randomly chosen between these two min. and max. values\\
  \hline
  $V_1^{(min)}$, $V_2^{(min)}$, $V_3^{(min)}$  & 8, 4, 1 & Minimum inherited assets of three income groups\\
  \hline
  $V_1^{(max)}$, $V_2^{(max)}$, $V_3^{(max)}$  & 10, 7, 3 & Maximum inherited assets of three income groups. Inherited asset of an agent is randomly chosen between these two min. and max. values\\
  \hline
  $W_1^{(min)}$, $W_2^{(min)}$, $W_3^{(min)}$  & 1, 3, 5 & Minimum welfare received by three income groups\\
  \hline
  $W_1^{(max)}$, $W_2^{(max)}$, $W_3^{(max)}$  & 2, 5, 7 & Maximum welfare received by three income groups. Welfare received by an agent is randomly chosen between these two min. and max. values\\
  \hline
  $\mu^{(crd)}_i$  & 10 & Assumed same for all agents \\
  \hline
  $\eta^{(rpy)}_i$  & 0.01 & Assumed same for all agents \\
  \hline
  $\rho^{(inv)}_i$  & 0.50 & Assumed same for all agents\\
  \hline
  $\rho^{(roi)}_i$  & 0.15 & Assumed same for all agents  \\
  \hline
  $\alpha^{(exp)}_{i,j}$  & 0.15 & Assumed same for all agents \\
  \hline
  $\alpha^{(env)}_{i,j}$  & 0.025 & Assumed same for all agents\\
  \hline
  $\alpha^{(tax)}_{i,j}$  & 1 & Assumed same for all agents\\
  \hline
  $\alpha^{(inc)}_{i,j}$  & 1 & Assumed same for all agents\\
  \hline
  $\alpha^{(wlf)}_{i,j}$  & 1 & Assumed same for all agents\\
  \hline
  $T$ & 300 & Simulation duration \\
  \hline
  $A_{i,j}$ & 10 & Assumed same for all agents\\
  \hline
  \end{tabular}
  \label{table_model_parameters}
\end{table}
\subsection{Impact of Expense}\label{section_result_exp}
\begin{figure}[ht]
\centering
\includegraphics[width=\columnwidth]{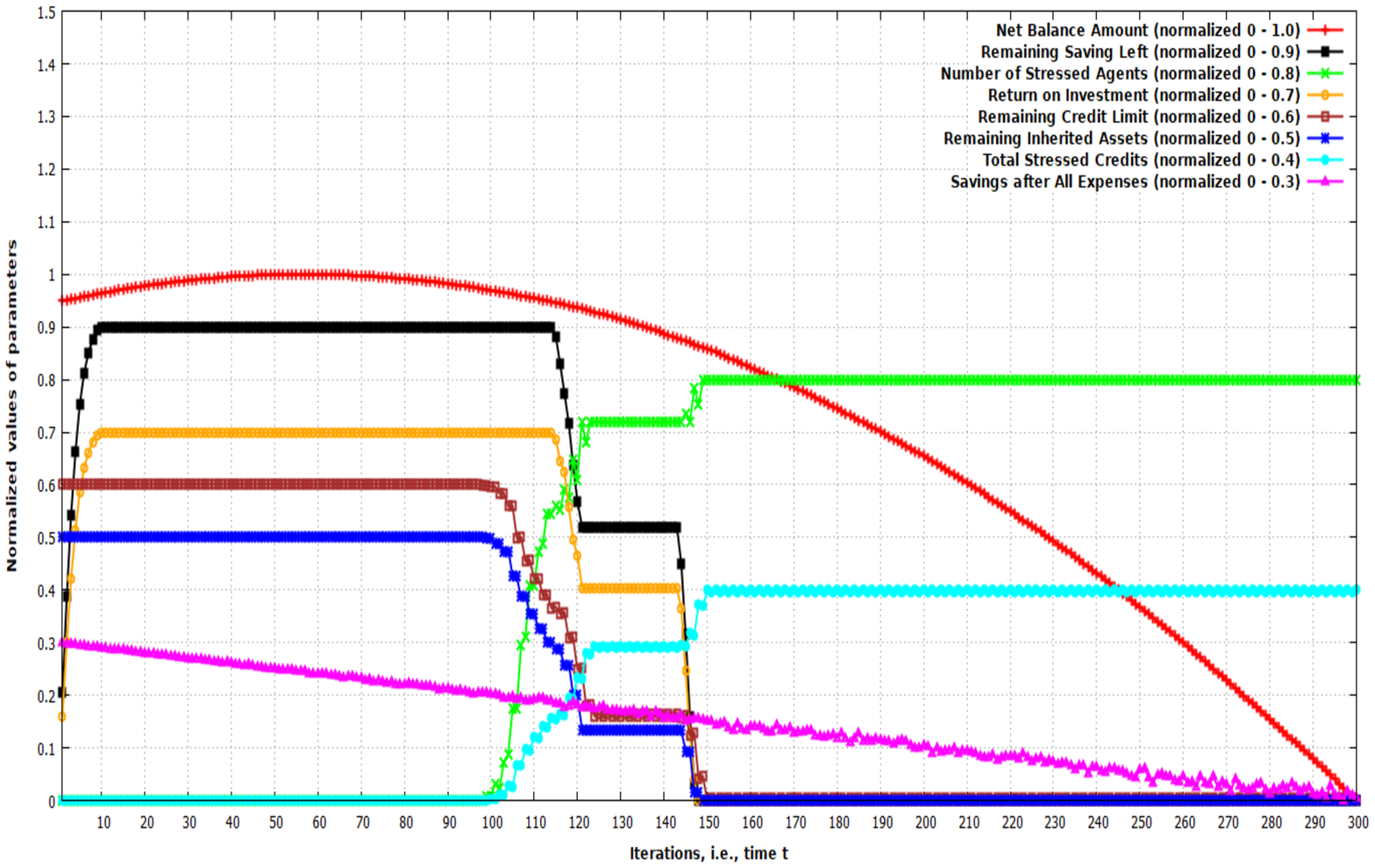}
\caption{Emergent behaviour of various parameters with increase in expense}
\label{fig_plot_exp_inkscape}
\end{figure}

Fig. \ref{fig_plot_exp_inkscape} shows the impact of increase in expense on various parameters and their emergent behaviours. In each iteration, the expense of each agent is increased by 15\% of its previous value, i.e.,  $\alpha^{(exp)} = 0.15$.
Even though the expenses increase, parameters RSL and ROI show improvements initial phase till iteration 10, i.e., $t=10$, and then stagnates, though SAE shows a downward trend continuously. NBA shows an increase till iteration $t=80$ and then starts declining with increased expenses. RIA and RCL still remains untouched, and NSA and TSC remain at their minimum values till $t=100$. However, there after the problem starts where agents start to meet their expenses through credits and RIA as both starts declining. A major deteriorating situation is observed around iterations $t=120$ where sharp decline in economic condition. Then, a short stable phase is observed. However, a complete collapse is observed round $t=150$. Thus, the model shows agents' deteriorating condition with increased expenses. One key thing is that the transitions can be sharp (first one from 100 to 122 and the second one from 145 to 150) implying that there are thresholds when reached can cause major economic problems for the agents within a short time. Also, the transitions get sharper over time.

\begin{figure}[ht]
\centering
\includegraphics[width=\columnwidth]{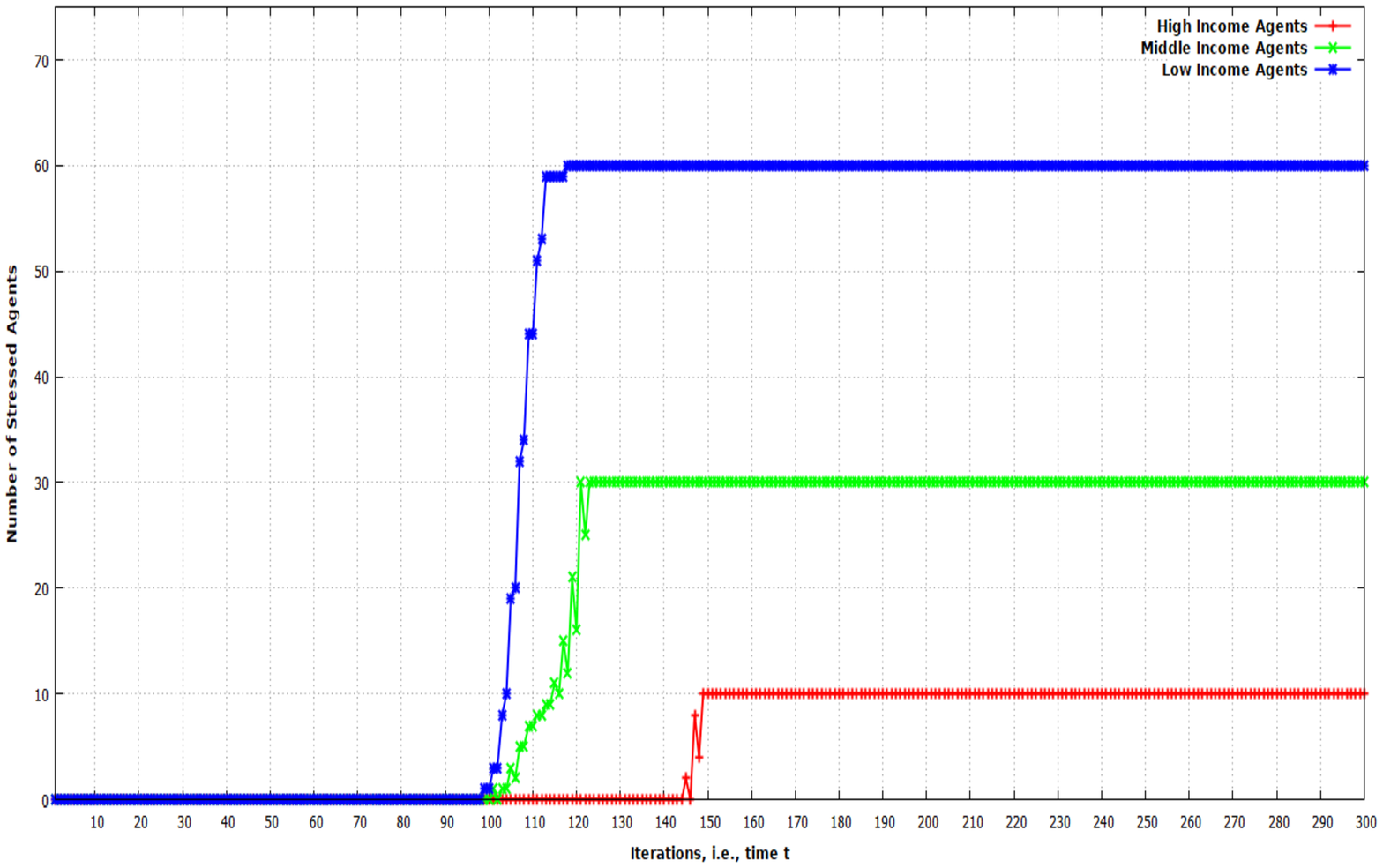}
\caption{Number of stressed agents under three income categories with increase in expense}
\label{fig_plot_exp_stressed_agents_inkscape}
\end{figure}

Fig. \ref{fig_plot_exp_stressed_agents_inkscape} shows the behaviour of stressed agents from three income groups with increase in expenses. It can be observed that as expense increases the number of stressed agents among the lower income group (blue line) gets impacted first and shows a sharp transition from minimum to maximum value from $t=98$ to $t=118$. The middle income group comes next with a longer phase transition from minimum to maximum from $t=100$ to $t=122$. The high income group is impacted last at a much later stage but still having a sharp transition from $t=145$ to $t=150$.
\subsection{Impact of Expense and Environment}\label{section_result_exp_env}
\begin{figure}[ht]
\centering
\includegraphics[width=\columnwidth]{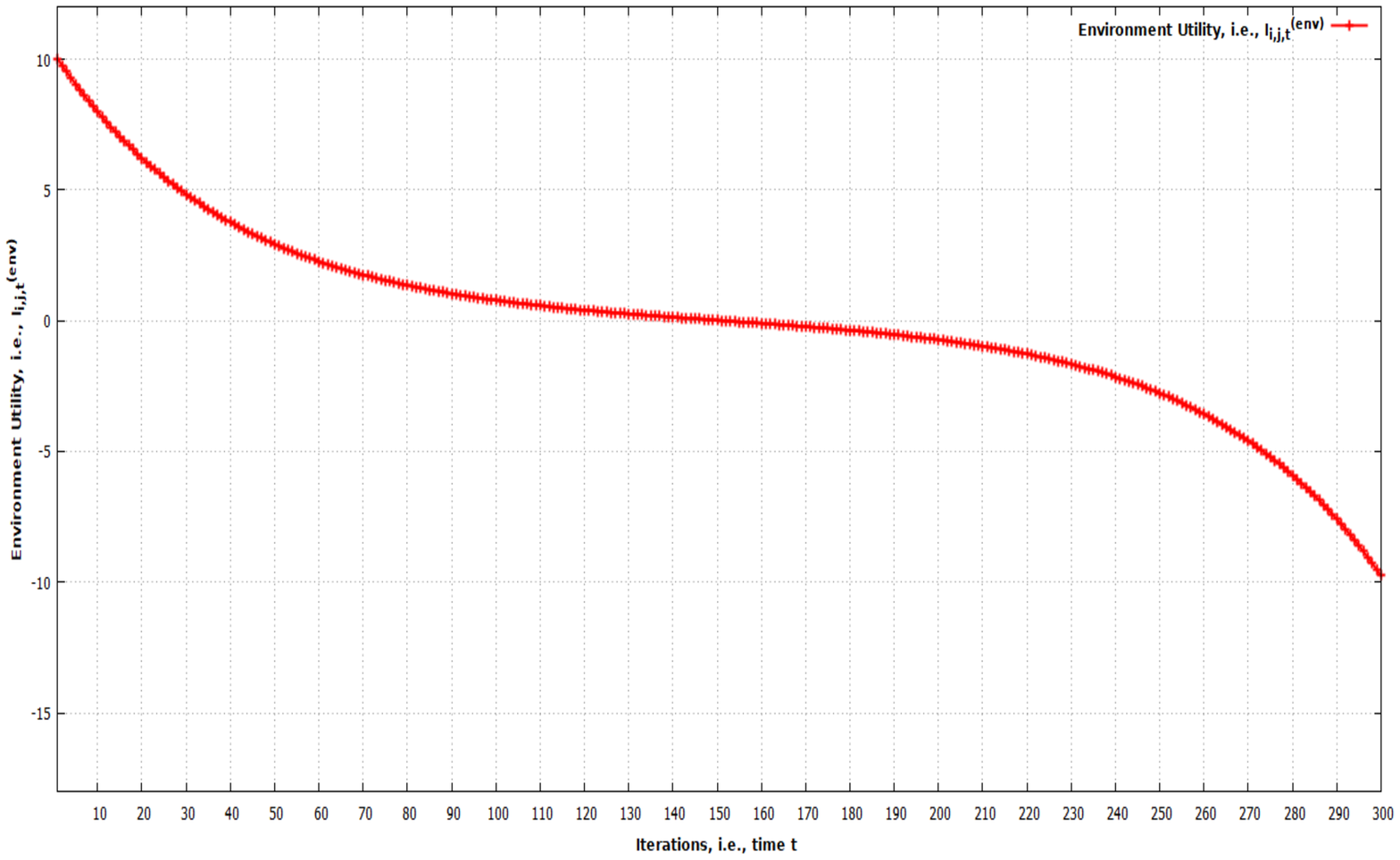}
\caption{Environment utility vs. time}
\label{fig_plot_exp_env_utility_inkscape}
\end{figure}

Fig. \ref{fig_plot_exp_env_utility_inkscape} shows the behaviour of environment utility, $I^{(env)}_{i,j,t}$ over time $t$. The values of $I^{(env)}_{i,j,t}$, $A_{i,j}$ and $\alpha^{(env)}_{i,j}$  are assumed same for all the agents to reduce complexity.

\begin{figure}[ht]
\centering
\includegraphics[width=\columnwidth]{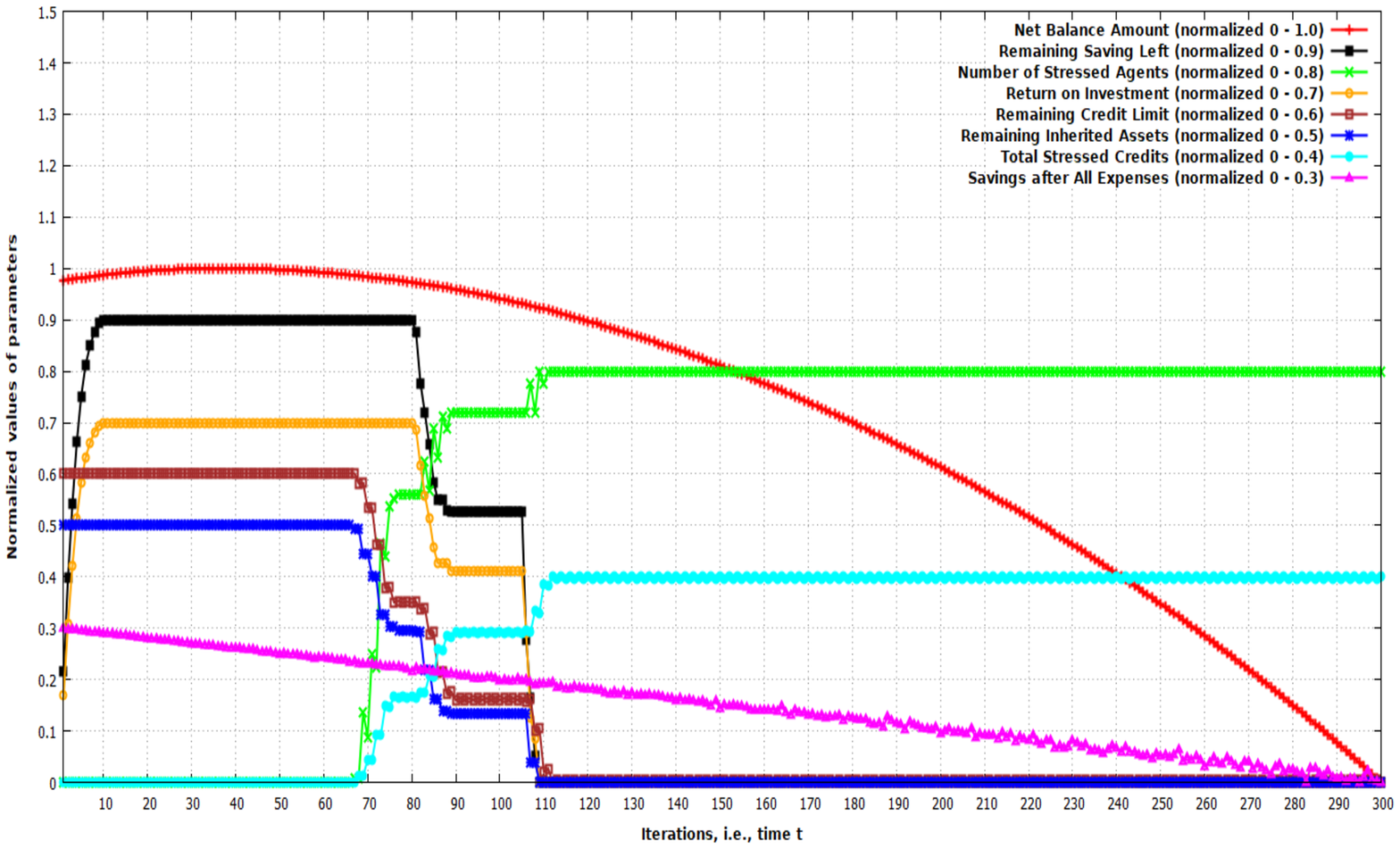}
\caption{Emergent behaviour of various parameters with increase in expense and decreasing environment utility}
\label{fig_plot_exp_env_inkscape}
\end{figure}

The emergent behaviour of various parameters with increase in expense and decrease in environment utility is plotted in Fig. \ref{fig_plot_exp_env_inkscape}. Though, the patterns are more or less similar to Fig. \ref{fig_plot_exp_inkscape} the transitions in Fig. \ref{fig_plot_exp_env_inkscape} occur much earlier between $t=67$ and $t=112$ as compared to the former. Also, the transitions are much more sharper compared to the previous case. The first transition occurs between $t=67$ and $t=90$ and the second one much sharper between $t=108$ and $t=112$.

The behaviour of the stressed agents from the three income groups also show transitions much earlier compared to Fig. \ref{fig_plot_exp_stressed_agents_inkscape}. The agents from lower income group shows sharper transition between $t=68$ and $t=78$ and those from other two groups remain the same between $t=68$ and $t=88$ and between $t=103$ and $t=108$ respectively.
\begin{figure}[ht]
\centering
\includegraphics[width=\columnwidth]{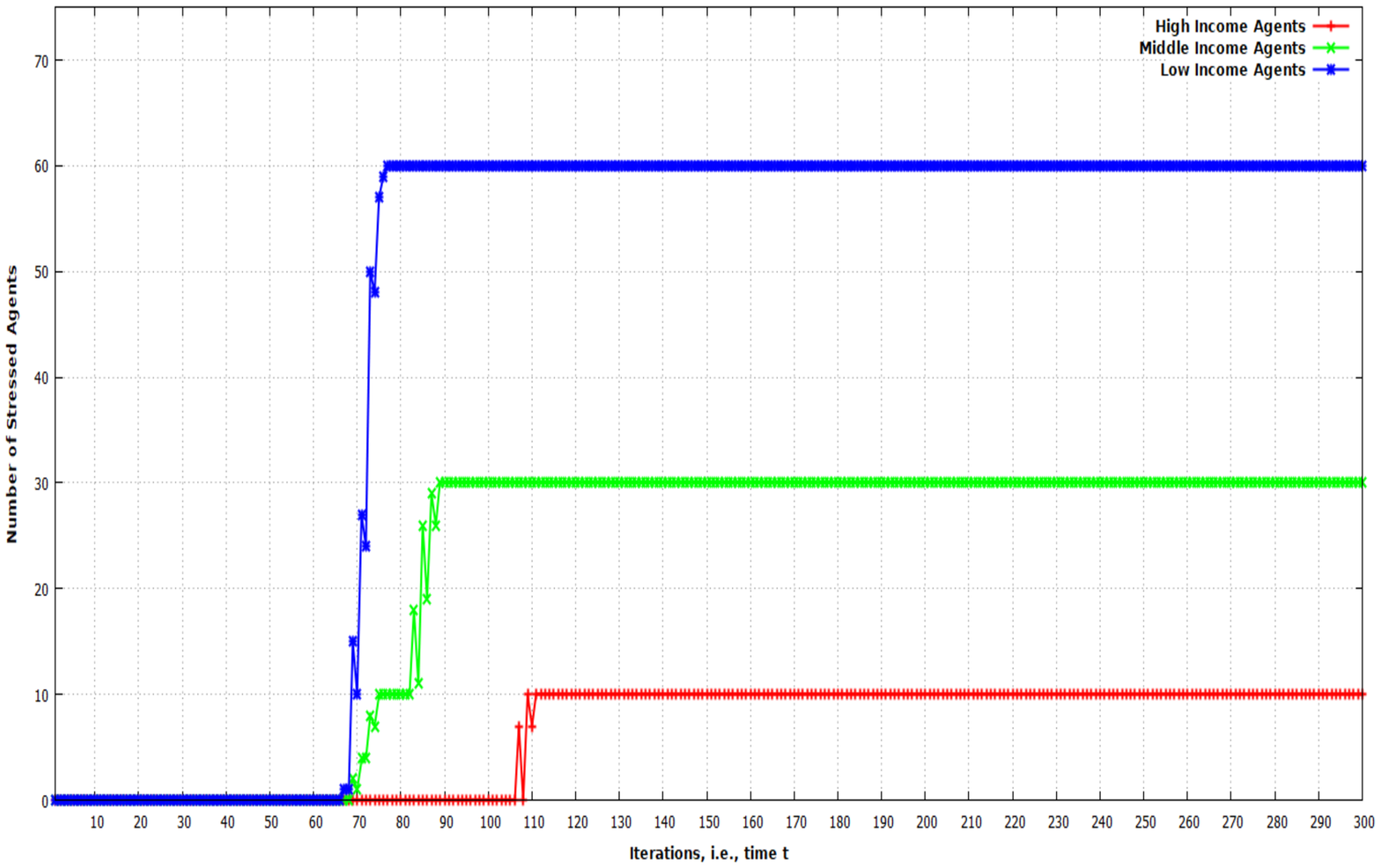}
\caption{Number of stressed agents under three income categories with increase in expense and decrease in environment utility}
\label{fig_plot_exp_env_stressed_agents_inkscape}
\end{figure}

Thus, the model shows that increase in expense and decrease in environment utility lead to much higher economic distress compared to the former situation. The lower income group gets stressed much earlier.
\subsection{Impact of Expense, Environment and Taxation}\label{section_result_exp_env_tax}
\begin{figure}[ht]
\centering
\includegraphics[width=\columnwidth]{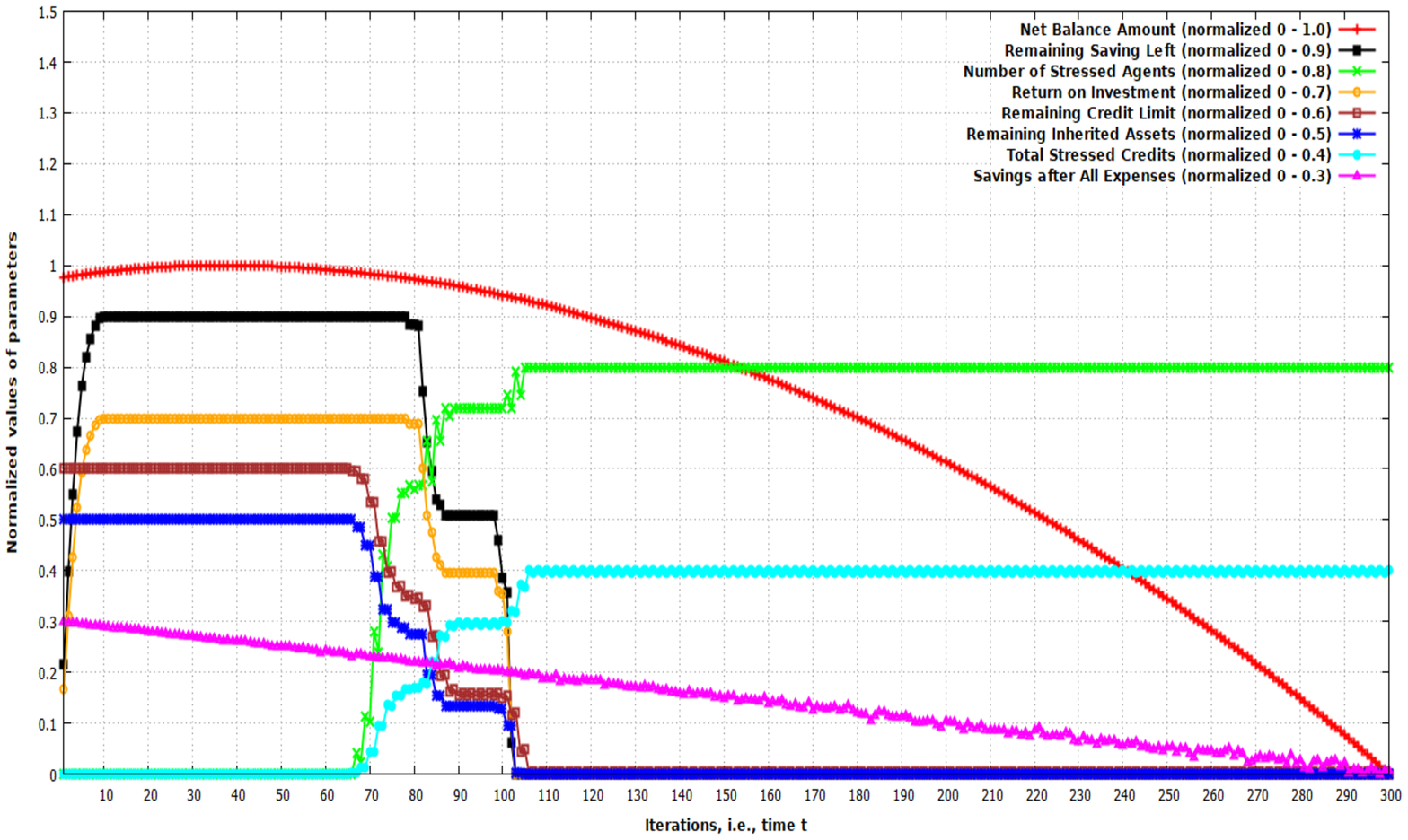}
\caption{Emergent behaviour of various parameters with increase in expense , decrease in environment utility and increase in taxes}
\label{fig_plot_exp_env_tax_inkscape}
\end{figure}

Fig. \ref{fig_plot_exp_env_tax_inkscape} shows the consequence of the different parameters when taxes increase along with increase in expense and decrease in environment utility for the agents. Though, the behaviour of the parameters remain the same as previous two scenarios, the transitions further shrinks in time, implying increased economic distress from $t=66$ to $t=89$ and then again $t=99$ from $t=105$.

The behaviour of the stressed agents of the three income groups shows faster transitions. Lower income groups shows a steeper curve from $t=65$ to $t=79$. Now, the middle income group also start showing steeper transitions between $t=70$ and $t=88$  as compared to the previous two cases. The high income group remains the same in its behaviour with transition between $t=100$ and $t=105$.
\begin{figure}[ht]
\centering
\includegraphics[width=\columnwidth]{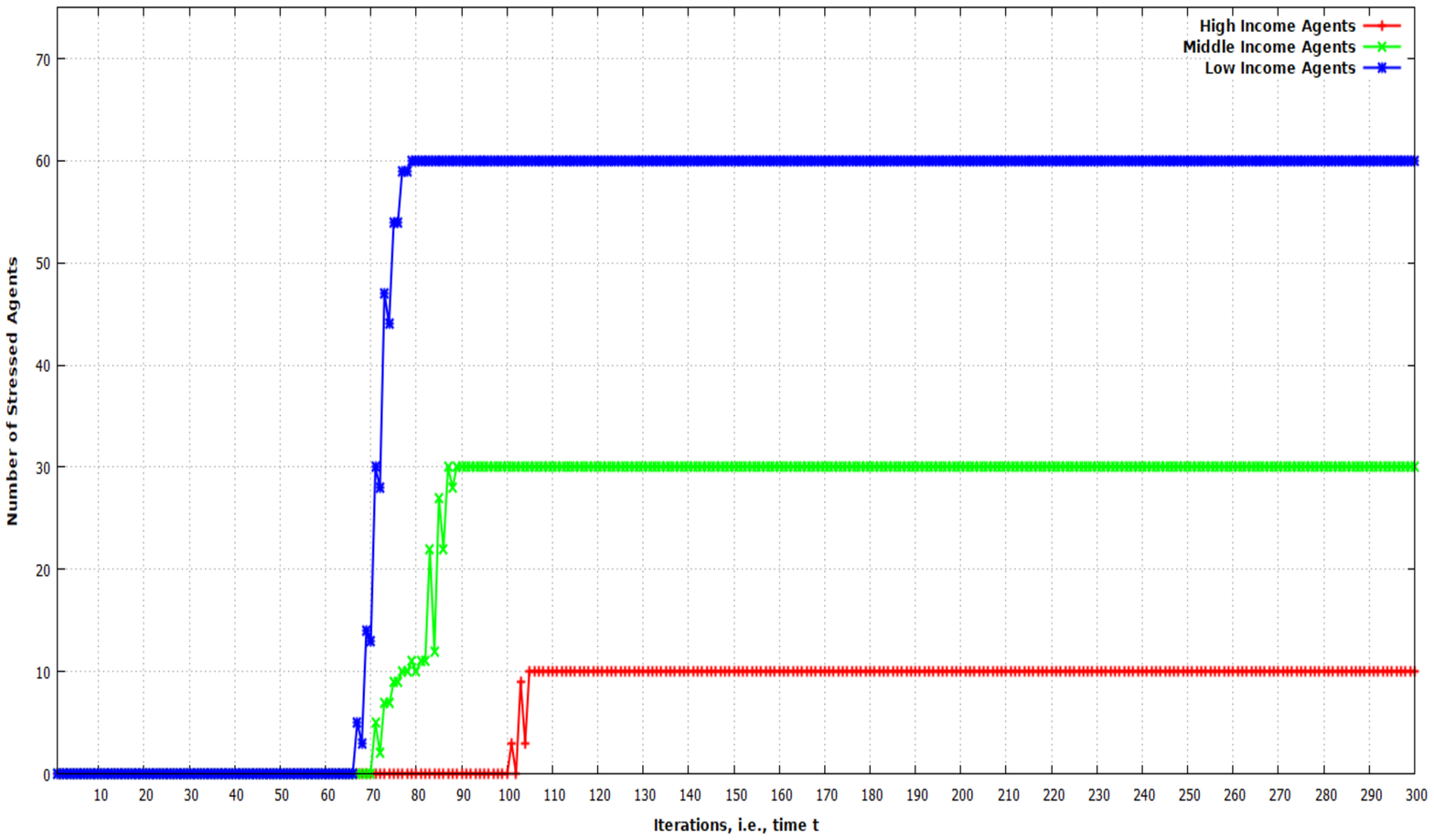}
\caption{Number of stressed agents under three income categories with increase in expense, decrease in environment utility and increase in taxes}
\label{fig_plot_exp_env_tax_stressed_agents_inkscape}
\end{figure}

Thus, the model shows decrease in environment utility and increases in expenses and taxes lead to quicker degradation of economic conditions of agents.
\subsection{Impact of Expense, Environment, Taxation and Welfare}\label{section_result_exp_env_tax}
\begin{figure}[ht]
\centering
\includegraphics[width=\columnwidth]{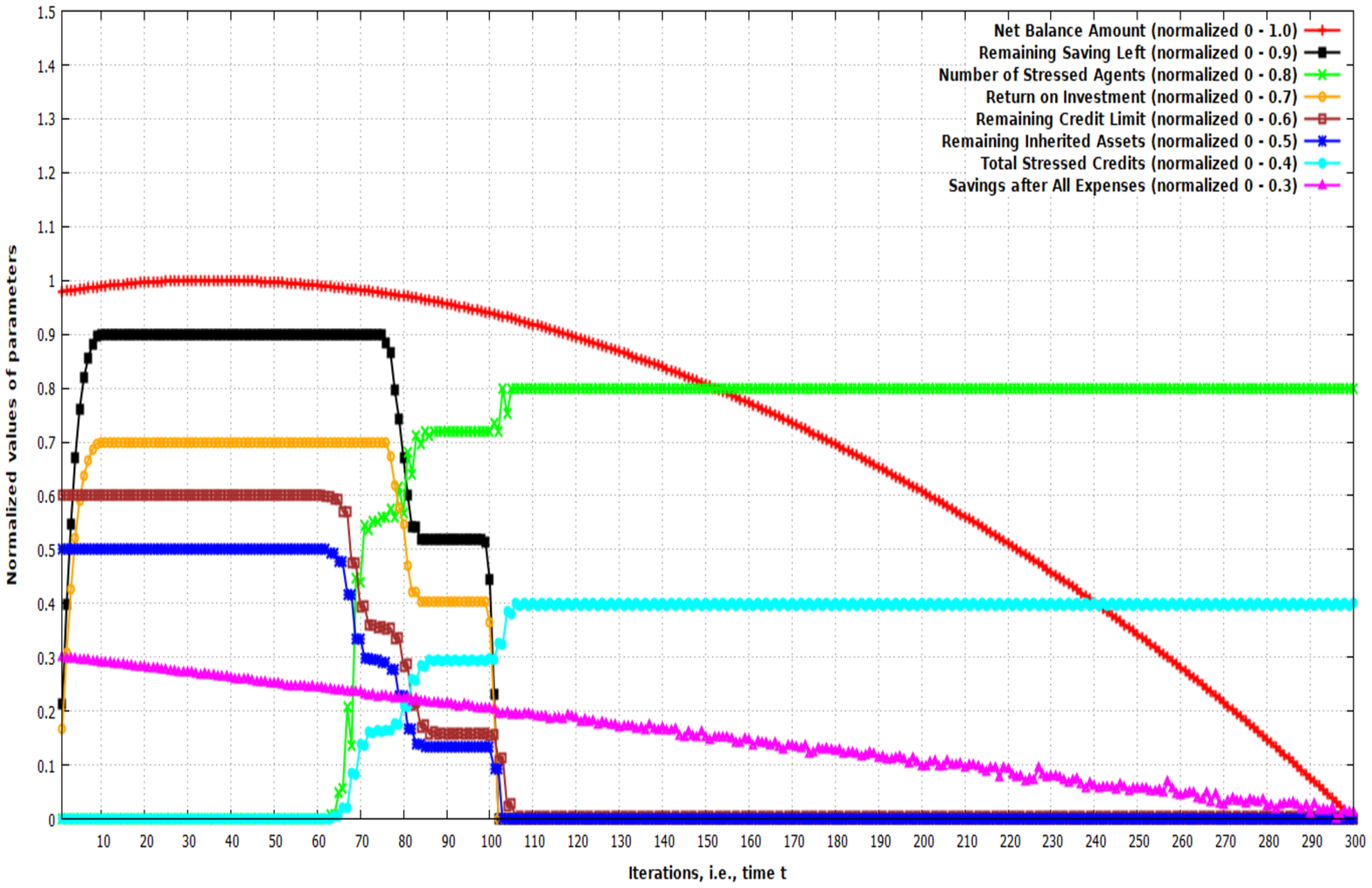}
\caption{Emergent behaviour of various parameters with increase in expense, decrease in environment utility, increase in taxes and decrease in public welfare received}
\label{fig_plot_exp_env_tax_welfare_inkscape}
\end{figure}

Fig. \ref{fig_plot_exp_env_tax_welfare_inkscape} shows the behaviour of all the parameters with increase in expense, decrease in environment utility, increase in tax and decrease in public welfare received. It can be observed that the transitions occur much quicker starting at $t=62$ till $t=84$ and collapsing at $t=99$ till $t=105$.

\begin{figure}[ht]
\centering
\includegraphics[width=\columnwidth]{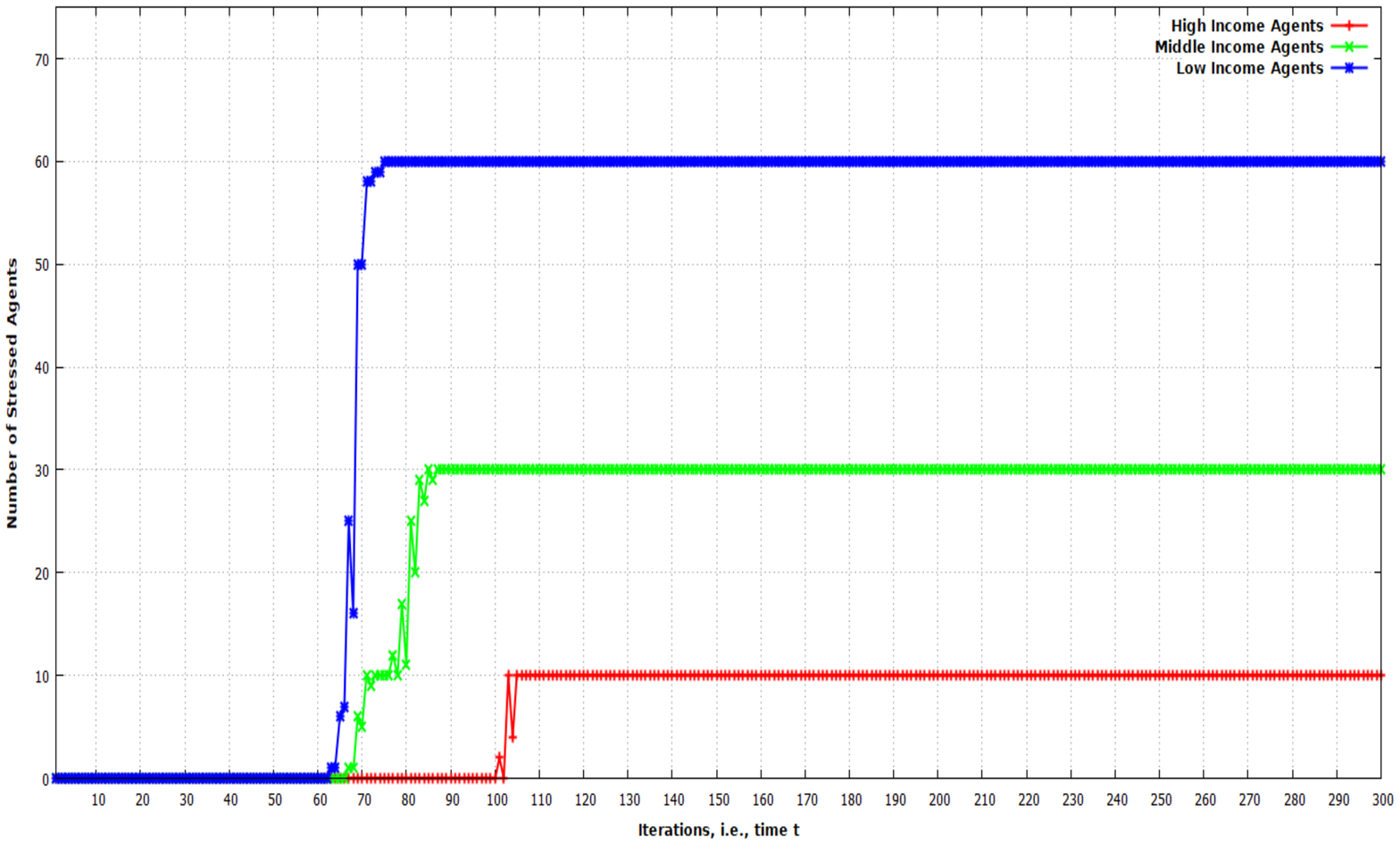}
\caption{Number of stressed agents under three income categories with increase in expense, decrease in environment utility and increase in taxes and decrease in public welfare utility}
\label{fig_plot_exp_env_tax_welfare_stressed_agents_inkscape}
\end{figure}

The stressed agents graph in Fig. \ref{fig_plot_exp_env_tax_welfare_stressed_agents_inkscape} also shows further sharper transition compared to all the previous three situations. Low income group show increase from $t=63$ to $t=75$, middle income from $t=69$ to $t=87$ and high income from $t=99$ to $t=104$. Thus, this scenario leads to quicker deterioration of economic conditions of the agents.
\subsection{Impact of Expense, Environment, Taxation,  Welfare and Income}\label{section_result_exp_env_tax_welfare_income}
Two scenarios are considered for reduced income of agents in addition to other factors in the previous scenarios. In the first case, agents reduce their expenses relative to their lower incomes. However in the second case, agents keep their expenses at their initial levels at $t=0$.
\subsubsection{Expense is based on current income $I^{(inc)}_{i,j,t}$}
\begin{figure}[ht]
\centering
\includegraphics[width=\columnwidth]{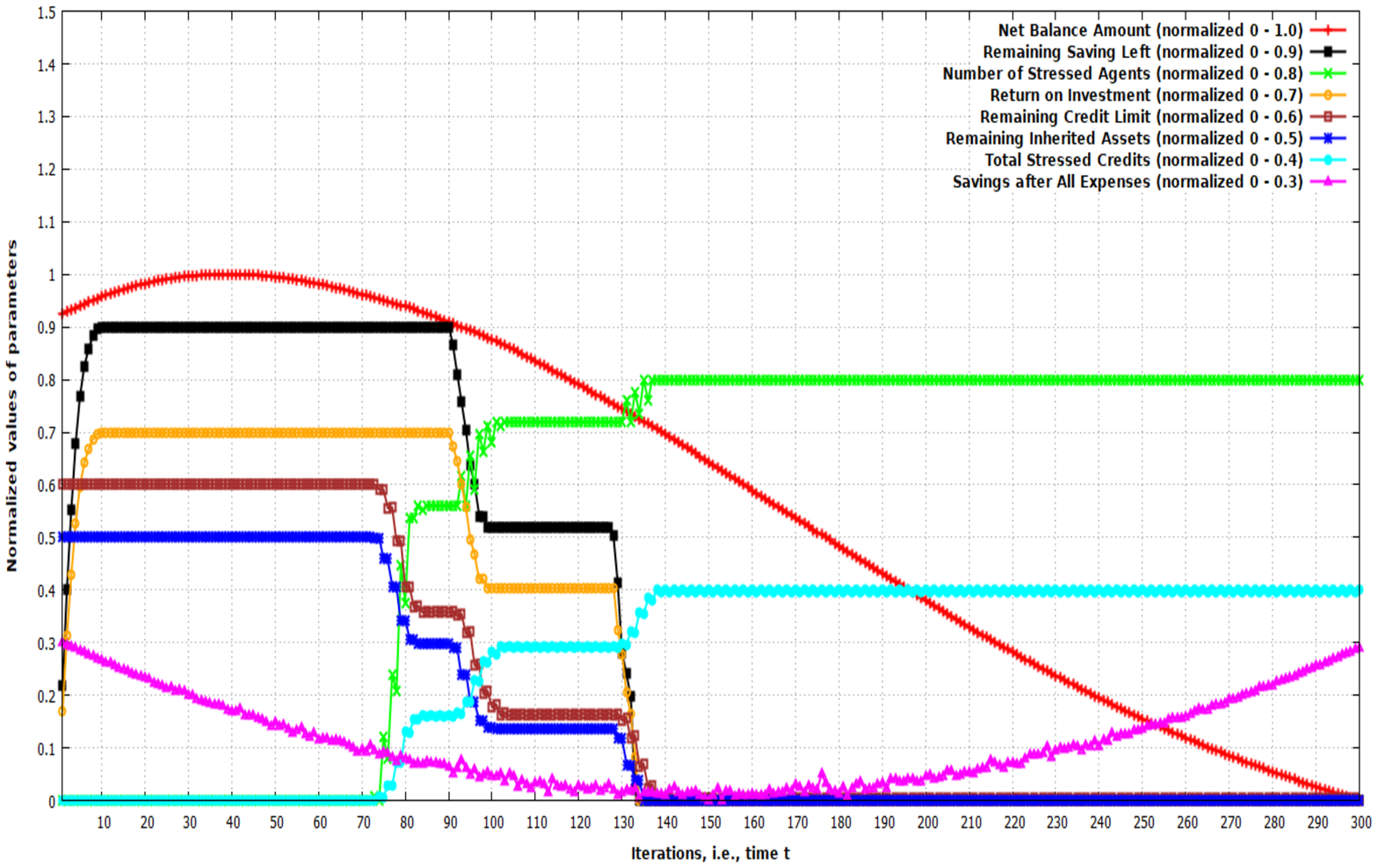}
\caption{Emergent behaviour of various parameters with increase in expense, decrease in environment utility, increase in taxes, decrease in public welfare received and decrease in income}
\label{fig_plot_exp_env_tax_welfare_income_inkscape}
\end{figure}

Fig. \ref{fig_plot_exp_env_tax_welfare_income_inkscape} shows the behaviour of parameters when increases in expense are fractions of present incomes of agents. Two things come into play in this scenario. Firstly, as income decreases the expenses of the agents also reduce. The increase in expense over time is relative to this reduced value. Secondly, when income decreases the direct taxes on income also comes down. It can be seen that the economic distress starts a bit late compared to the previous cases at $t=73$ and collapses completely at $t=137$ within a duration of 64. First transition happens between $t=73$ and $t=98$ and second one happens between $t=123$ and $t=137$. Also, SAEs increase due to decrease in taxes. However, these do not help stop the collapse due to other deteriorating factors.

\begin{figure}[ht]
\centering
\includegraphics[width=\columnwidth]{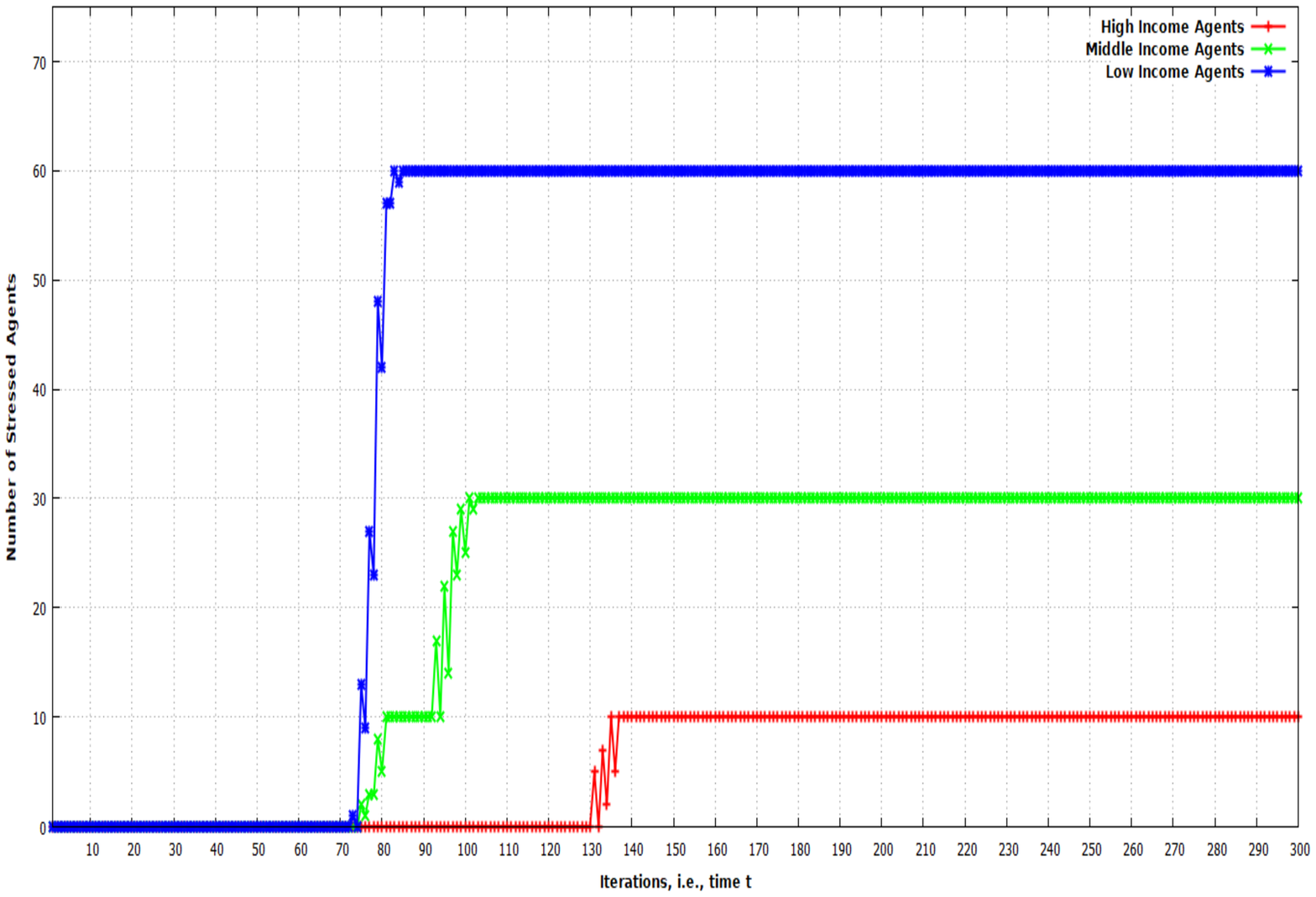}
\caption{Number of stressed agents under three income categories with increase in expense, decrease in environment utility and increase in taxes, decrease in public welfare received and decrease in income}
\label{fig_plot_exp_env_tax_welfare_income_stressed_agents_inkscape}
\end{figure}

Low income groups deteriorates faster due to reduced income (Fig. \ref{fig_plot_exp_env_tax_welfare_income_stressed_agents_inkscape}). The middle and high income groups hold their ground for a longer duration before the final collapse.
\subsubsection{Expense is based on $I^{(inc)}_{i,j,0}$}
\begin{figure}[ht]
\centering
\includegraphics[width=\columnwidth]{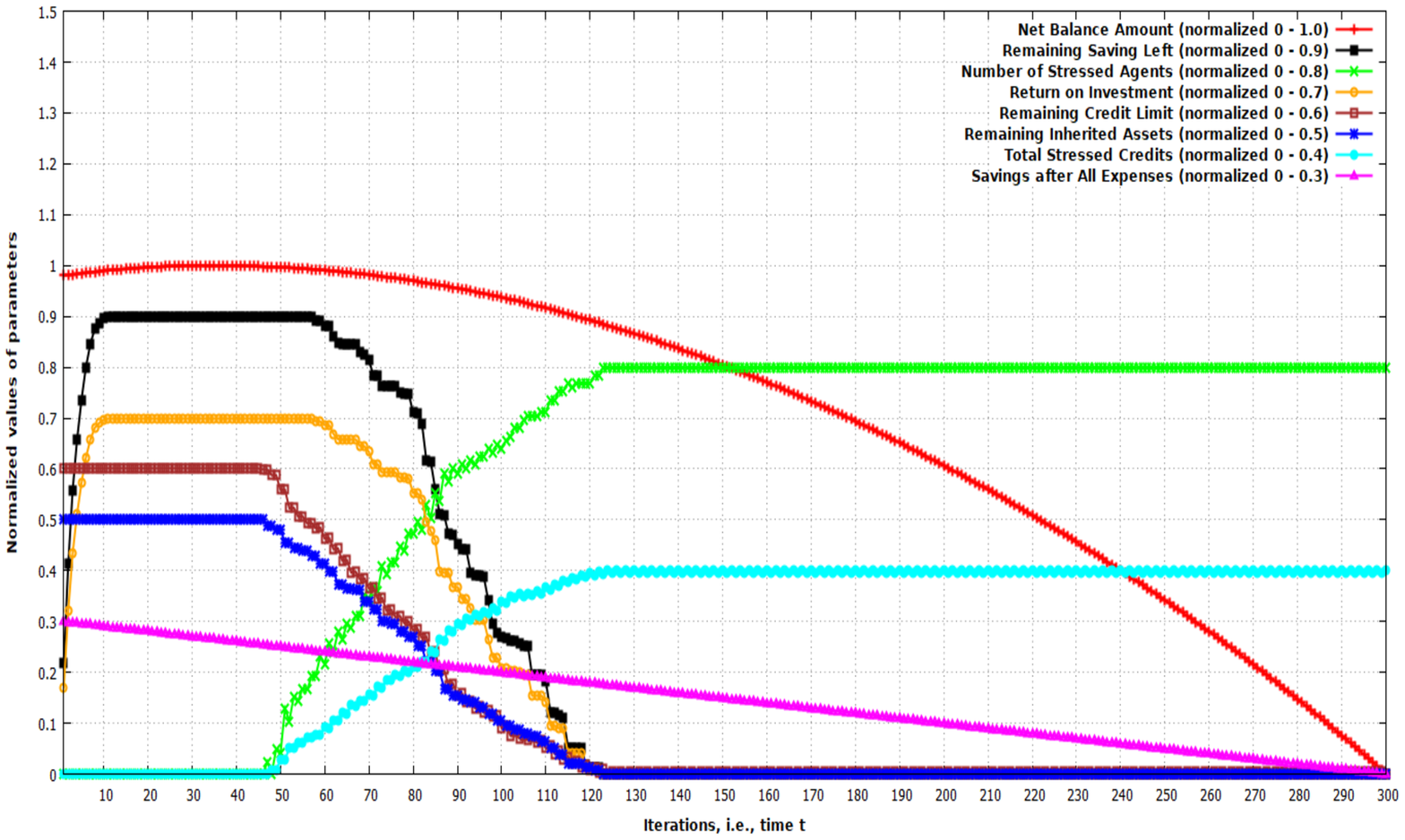}
\caption{Emergent behaviour of various parameters with increase in (original) expense (fixed), decrease in environment utility, increase in taxes, decrease in public welfare received and decrease in income}
\label{fig_plot_exp_fixed_env_tax_welfare_income_inkscape}
\end{figure}

Fig. \ref{fig_plot_exp_fixed_env_tax_welfare_income_inkscape} shows the behaviour when expenses remain the same as at $t=0$ increase over time. Results show that the distresses start much early at $t=47$ and drag on till $t=123$ due to decreases in direct taxes. Surprisingly, there is only one declining trend unlike the previous cases where there is a stable phase between the two sharp transitions. Thus, decrease in income can bring distress much earlier with a continuous downward trend.

\begin{figure}[ht]
\centering
\includegraphics[width=\columnwidth]{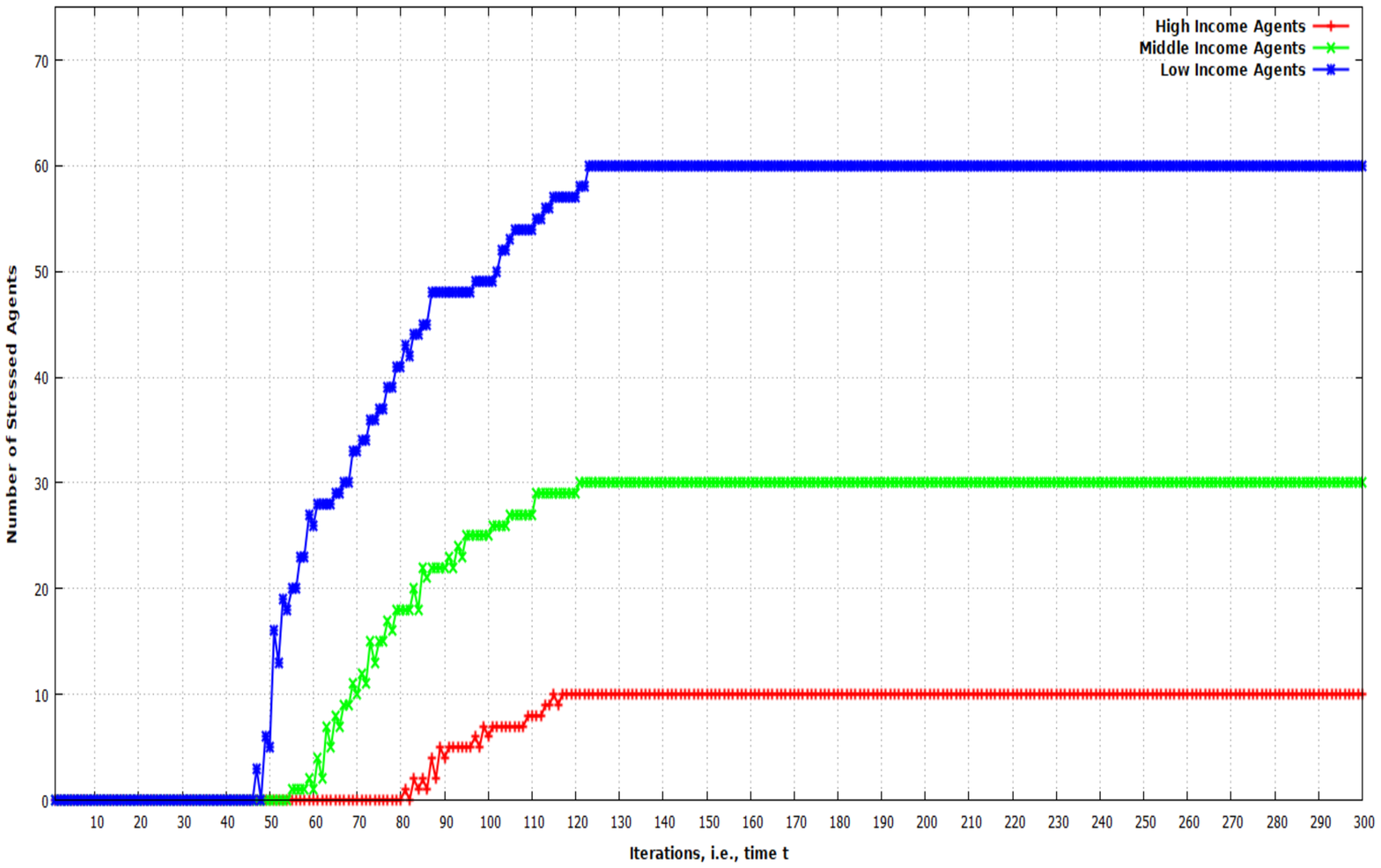}
\caption{Number of stressed agents under three income categories with increase in expense, decrease in environment utility and increase in taxes, decrease in public welfare received and decrease in income}
\label{fig_plot_exp_fixed_env_tax_welfare_income_stressed_agents_inkscape}
\end{figure}

Another interesting fact is that the agents from all three income groups reach their peak values almost at the same time at $t=120$ (Fig. \ref{fig_plot_exp_fixed_env_tax_welfare_income_stressed_agents_inkscape}) even though distresses start much earlier at different times and are gradual compared to previous scenarios.
\section{Discussion}
Generally, following observations can be made of agents with progressively difficult economic conditions in the above five scenarios. Firstly, there are two transitions which are sharp implying that the economic collapse of agents can be sudden. Secondly, the second transition is sharper than the first which implies that the complete collapse can be faster. Thirdly, with increased economic burdens the transitions happen progressively faster and sharper, except for the case of decrease in income when it is delayed because of lower direct taxes to be paid by the agents. Fourthly, as expected, low income agents tend to collapse much faster, followed by the middle income group and finally the high income group. However, it is just a matter of time before all the agents are affected.
\section{Conclusion}
Post-pandemic world is seeing many challenges. These include high inflation, low growth, high debt, collapse of economies, political instability, job losses, lowering of income in addition to damages caused natural disasters and existing inequalities. Efforts are being made to mitigate these challenges at various levels. Research works have focussed on policies on specific scenarios, use cases, relationships between couple of sectors. However, understanding of actual impact of policies on individual agents and changing dynamics using a computational model has been largely unexplored. This paper explores progressive deteriorating conditions of increase in expense, degrading environmental utility, increase in taxation, decrease in welfare and lowering of income on agents with options to inherited properties, credits and return on investments using an agent based model. The paper also defines several parameters related to economic health of the group of agents in terms savings, ROI, credits, etc.  Results indicate that collapse of agents' economic condition can happen sharply for all income groups with increasing difficult conditions.

Future work will focus on more dynamic scenarios with different rate of change of parameters, such as, expense, income, tax, welfare, etc.

\bibliographystyle{ACM-Reference-Format}
\bibliography{ref}

\end{document}